\documentclass[pre,aps,twocolumn,superscriptaddress,showpacs]{revtex4}

\usepackage{epsfig,amsmath,amssymb}
\newcommand{\rv}{{\bf r}}
\newcommand{\xv}{{\bf x}}
\newcommand{\Tr}{{\rm Tr}\,}

\newcommand{\J}{{\bf J}}
\newcommand{\N}{{\sf N}}
\newcommand{\F}{{\sf F}}
\newcommand{\dE}{b}

\begin{document}

\title{Statics and dynamics of inhomogeneous liquids via the internal-energy functional}

\author{Matthias Schmidt}
\affiliation{Theoretische Physik II, Physikalisches Institut, 
  Universit{\"a}t Bayreuth, D-95440 Bayreuth, Germany}

\date{20 July 2011, resubmitted: 17 September 2011, to appear in Phys. Rev. E}

\begin{abstract}
   We give a variational formulation of classical statistical
   mechanics where the one-body density and the local entropy
   distribution constitute the trial fields. Using Levy's constrained
   search method it is shown that the grand potential is a functional
   of both distributions, that it is minimal in equilibrium, and that
   the minimizing fields are those at equilibrium. The functional
   splits into a sum of entropic, external energetic and internal
   energetic contributions. Several common approximate Helmholtz free
   energy density functionals, such as the Rosenfeld fundamental
   measure theory for hard sphere mixtures, are transformed to
   internal energy functionals. The variational derivatives of the
   internal energy functional are used to generalize dynamical density
   functional theory to include the dynamics of the microscopic
   entropy distribution, as is relevant for studying heat transport
   and thermal diffusion.
\end{abstract}


\maketitle

\section{Introduction}

The study of classical many-body systems in equilibrium is often based
on the grand potential $\Omega_0$ as a function of its natural
variables, which for a one-component system are the chemical potential
$\mu$, the temperature $T$ and the volume $V$. The mean number of
particles, $N_0$, is then obtained as a partial derivative $\partial
\Omega_0/\partial \mu = -N_0$, while keeping $T$ and $V$ fixed, and
the mean particle density is simply $N_0/V$. When the Hamiltonian
contains a contribution due to an external potential, $v(\rv)$, where
$\rv$ is the position coordinate, then the density is in general no
longer uniform, but becomes position-dependent, hence $\rho_0(\rv)$,
where $\rho_0(\rv)d\rv$ gives the mean number of particles in a volume
element $d\rv$. Here the difference $\mu-v(\rv)$ and $\rho_0(\rv)$
play the role of conjugate thermodynamic fields, and for convenience
one often defines (formally) a position-dependent chemical potential
$\mu(\rv)=\mu-v(\rv)$. The one-body density distribution can then be
obtained as a functional derivative $\delta \Omega_0/\delta \mu(\rv) =
-\rho_0(\rv)$.

Density functional theory (DFT) \cite{evans79} amounts to generalizing
this concept to a functional dependence of the grand potential on the
one-body density distribution, i.e., going from $\Omega_0(\mu,T,V)$ to
a functional $\Omega([\rho],\mu,T,V)$, where $\rho(\rv)$ is a trial
field.  The variational principle of DFT \cite{mermin65,evans79}
states that for given thermodynamic state (i.e.\ fixed values of $T$,
$\mu$, and $V$) the density distribution that minimizes $\Omega$ is
the physically realized equilibrium density $\rho_0(\rv)$. The
non-trivial (additive) contribution to~$\Omega$ is the Helmholtz free
energy functional $F([\rho],T,V)$, which is independent of $\mu$ and
generalizes the equilibrium Helmholtz free energy $F_0(N_0,T,V)$ to a
functional dependence on the trial density $\rho(\rv)$. Inserting the
equilibrium density into the functional yields the equilibrium free
energy, i.e., $F_0(N_0,T,V)=F([\rho_0],T,V)$, where $N_0=\int d\rv
\rho_0(\rv)$. There is a significant body of literature on application
of this framework to a wide variety of interesting many-body phenomena
in liquids and solids \cite{evans90,roth10review,lutsko10review}.
Both conceptually, and in practical DFT applications, the temperature
enters as a mere parameter, often in the form of ``thermal energy''
$k_BT$, where $k_B$ is the Boltzmann constant. Clearly this situation
is very different from the sophisticated treatment that the chemical
potential received via introduction of $\mu(\rv)$ and its conjugate
field $\rho(\rv)$. One might justify this by the fact that
$\mu(\rv)\neq \rm const$ leads to a well-defined equilibrium when
$v(\rv)$ acts on the system, whereas considering inhomogeneous
temperature distributions reeks of non-equilibrium.

In thermodynamics, one can proceed and Legendre transform to the
internal energy $E_0(N_0,S,V)$, where the entropy $S$ is an extensive
state variable, conjugate to temperature $T$. The latter is recovered
from $T=\partial E_0/\partial S$ at $N_0,V=\rm const$. Were one to
generalize to an internal energy functional, one needed to introduce
(and define) an entropy distribution $s(\rv)$ that would ``localize''
(i.e. make dependent on position) the bulk entropy per unit volume,
$S/V$. This programme possesses several requirements in order to be
rigorous. i) A microscopic definition of the entropy distribution
$s(\rv)$ needs to be given.  ii) The grand potential functional needs
to depend on both the microscopic density and the microscopic entropy,
i.e., $\Omega([\rho,s],\mu,T,V)$. Its non-trivial contribution should
be an internal energy functional of both microscopic distributions,
i.e., $E([\rho,s],V)$. iii) The generalized grand potential functional
should be minimal at the equilibrium values $\rho_0(\rv)$ and
$s_0(\rv)$. Note that $T=\rm const$ will be associated in general with
a non-trivial spatial dependence $s_0(\rv)\neq\rm const$. This forms a
generalization of the simple parametric dependence on temperature in
conventional DFT to a proper Euler-Lagrange equation.

In the following such a framework is established. We formulate the
variational principle in Sec.\ \ref{SECvariationalPrinciple}.
Standard DFT approximation are converted to the internal energy
representation in Sec.~\ref{SECapproximations}. This includes internal
energy functionals for the ideal gas, hard spheres in the fundamental
measures approximation
\cite{rosenfeld89,kierlik90,roth10review,tarazona08review}, the
quadratic mean-field functional \cite{evans92}, etc. Based on the
continuity equations for particle density and internal energy density,
and inspired by the framework of linear irreversible thermodynamics,
in Sec.\ \ref{SECdynamics} we formulate a theory for diffusive
dynamics that corresponds to dynamical DFT (DDFT)
\cite{evans79,archer04ddft,marinibettolomarconi99}, but includes the
dynamics of the entropy current. Conclusions are given in
Sec.\ \ref{SECconclusions}.

\section{Variational principle}
\label{SECvariationalPrinciple}
We consider a classical system with $N$ particles and Hamiltonian
$H_N$. The equilibrium many-body probability distribution in the grand
ensemble is given by
\begin{align}
  f_0 = \Xi^{-1} \exp\left(-\frac{H_N-\mu N}{k_BT}\right).
  \label{EQf0}
\end{align}
Here the normalization constant is the grand partition sum
\begin{align}
  \Xi = \Tr \exp\left(-\frac{H_N-\mu N}{k_BT}\right),
\end{align}
with the (classical) trace being defined as
\begin{align}
  \Tr = \sum_{N=0}^\infty \frac{1}{h^{3N}N!}\int d\rv_1\ldots d\rv_N
   \int d{\bf p}_1 \ldots d{\bf p}_N,
\end{align}
where $h$ is Planck's constant, $\rv_i$ is the position coordinate and
${\bf p}_i$ is the momentum of particle $i=1,\ldots,N$.  Mermin's form
\cite{mermin65} for the grand potential as a functional of a trial
many-body distribution $f$ is
\begin{align}
  \Omega[f] = \Tr f \left(H_N - \mu N + k_BT \ln f \right).
  \label{EQomega_f}
\end{align}
Here $f$ is an arbitrary many-body distribution that is normalized,
i.e., that satisfies
\begin{align}
  \Tr f = 1.
\end{align}
Inserting the equilibrium distribution (\ref{EQf0}) into
(\ref{EQomega_f}) yields
\begin{align}
  \Omega[f_0] & = \Tr f_0 \left( H_N - \mu N + k_BT\ln f_0 \right)
  \label{EQomegaf0}\\
  & = \Tr f_0 \left[ H_N - \mu N - k_BT  \left(
      \ln\Xi + \frac{H_N-\mu N}{k_BT}
      \right) \right]\\
  & = -k_BT \ln\Xi\\
  & \equiv \Omega_0,
\end{align}
where $\Omega_0$ is the equilibrium grand potential. From the Gibbs
inequality it is straightforward to show \cite{evans79,hansen06} that
for any $f\neq f_0$ the inequality $\Omega[f] > \Omega[f_0]$ holds and
hence
\begin{align}
  \Omega_0 = \min_f \Tr f (H_N-\mu N+k_BT \ln f),
  \label{EQomegaFromMinf}
\end{align}

We use the conventional definition of the density operator
\cite{evans79,hansen06},
\begin{align}
  \hat \rho(\rv) = \sum_{i=1}^N \delta(\rv-\rv_i),
\end{align}
where $\delta(\cdot)$ is the (three-dimensional) Dirac distribution,
and express the one-body density distribution in equilibrium as the
average
\begin{align}
  \rho_0(\rv) = \Tr \hat\rho(\rv)f_0.
  \label{EQrho0}
\end{align}
We also define a position-dependent entropy density (per unit volume)
as
\begin{align}
  s_0(\rv) = -k_B\Tr \frac{\hat\rho(\rv)}{N}f_0\ln f_0.
  \label{EQs0}
\end{align}
Note that the integral $-T \int d\rv s_0(\rv) = \Tr k_BT f_0 \ln f_0$
equals the entropic contribution to the grand potential, cf.\ the last
term in Eq.~(\ref{EQomegaf0}).

We use Levy's constrained search method \cite{levy79,levy2010}, as
proved useful for classical systems \cite{dwandaru11levy}, and express
(\ref{EQomegaFromMinf}) as a two-stage minimization
\begin{align}
 \Omega_0 = \min_{\rho,s} \min_{f\to\rho,s} \Tr f (H_N-\mu N+k_BT\ln f),
  \label{EQomega0asDoubleMinization}
\end{align}
where the inner minimization is performed for all trial $f$ under the
constraint that these generate the given density distribution
$\rho(\rv)$ and the given local entropy distribution $s(\rv)$ via
\begin{align}
  \rho(\rv) &= \Tr \hat\rho(\rv) f, \label{EQfToRho}\\
  s(\rv) &= -k_B\Tr \frac{\hat\rho(\rv)}{N}f\ln f. \label{EQfTOs}
\end{align}
The relationships (\ref{EQfToRho}) and (\ref{EQfTOs}) are indicated as
$f\to\rho,s$ in the notation of (\ref{EQomega0asDoubleMinization}).

In the following we restrict ourselves to Hamiltonians that consist of
kinetic energy and internal and external contributions to the
potential energy, i.e., that are of the form
\begin{align}
 H_N = \sum_{i=1}^N \frac{p_i^2}{2m} + U(\rv_1,\ldots,\rv_N) + 
 \sum_{i=1}^N v(\rv_i),
\end{align}
where $p_i^2 = {\bf p}_i\cdot{\bf p}_i$, $m$ is the particle mass, $U$
is the interparticle interaction potential, and $v(\rv)$ is an
external potential that acts on the system.  Hence
(\ref{EQomega0asDoubleMinization}) is more explicitly
\begin{align}
  \Omega_0 &= \min_{\rho,s} \min_{f\to\rho,s} \Tr f 
   \left(\sum_{i=1}^N\frac{p_i^2}{2m} + U(\rv_1,\ldots,\rv_N) \right.
   \nonumber\\
   &\left.\qquad\qquad\qquad+\sum_{i=1}^N v(\rv_i)
   -\mu N+k_BT\ln f\right).
   \label{EQomega0asDoubleMinizationExplicit}
\end{align}
In the expression above several contributions can be written as space
integrals over averaged one-body quantities. First, the terms due to
the external and the chemical potential are
\begin{align}
  \Tr f \left(\sum_{i=1}^N v(\rv_i)-\mu N\right) =
  \int d\rv \rho(\rv) (v(\rv)-\mu),
  \label{EQinteriorExternalEnergy}
\end{align}
because $f\to\rho$ via (\ref{EQfToRho}). Furthermore, the last term in
(\ref{EQomega0asDoubleMinizationExplicit}) is
\begin{align}
 \Tr f k_B T \ln f = -T\int d\rv s(\rv),
 \label{EQinteriorEntropy}
\end{align}
because (\ref{EQfTOs}) implies that $f\to s$. Hence the terms
(\ref{EQinteriorExternalEnergy}) and (\ref{EQinteriorEntropy}) are
constants with respect to the inner minimization in
(\ref{EQomega0asDoubleMinization}). Hence we can separate them out and
arrive at
\begin{align}
  \Omega_0 = \min_{\rho,s}\left\{ E[\rho,s] +
  \int d\rv \left[\rho(\rv) (v(\rv)-\mu) - Ts(\rv) \right]
  \right\},
  \label{EQomega0decomposed}
\end{align}
where we have defined the internal energy as a functional of the
density and entropy distributions as
\begin{align}
  E[\rho,s] = \min_{f\to\rho,s} \left[f \left(\sum_{i=1}^N\frac{p_i^2}{2m} 
  + U(\rv_1,\ldots,\rv_N) \right)\right],
\end{align}
where, once more, the minimization (``search'' \cite{levy79}) is
constrained to all trial $f$ that generate the given $\rho(\rv)$ and
$s(\rv)$ via (\ref{EQfToRho}) and (\ref{EQfTOs}), respectively.  Here
and in the following we suppress the dependence on volume $V$ in the
notation.

The grand potential functional is the object inside of the
minimization in (\ref{EQomega0decomposed}), defined as
\begin{align}
  \Omega([\rho,s],\mu,T) = E[\rho,s] - T\int d\rv s(\rv) +
  \int d\rv\rho(\rv)(v(\rv)-\mu).
  \label{EQomegaDefinition}
\end{align}
Eq.~(\ref{EQomega0decomposed}) then becomes
\begin{align}
  \Omega_0 = \min_{\rho,s} \Omega([\rho,s],\mu,T),
\end{align}
which implies that the following functional derivatives vanish at
equilibrium
\begin{align}
  \left.\frac{\delta \Omega([\rho,s],\mu,T)}{\delta \rho(\rv)}\right|_{\rho_0,s_0} = 0
  \quad {\rm and} \quad
  \left.\frac{\delta \Omega([\rho,s],\mu,T)}{\delta s(\rv)}\right|_{\rho_0,s_0} = 0.
  \label{EQomegaDerivativesVanish}
\end{align}
The density and entropy distribution that satisfy
(\ref{EQomegaDerivativesVanish}) are indeed $\rho_0(\rv)$ and
$s_0(\rv)$, as can be seen from their definitions, (\ref{EQrho0}) and
(\ref{EQs0}), and the fact that $f_0$ minimizes $\Omega[f]$. This implies
that 
\begin{align}
  \Omega_0(\mu,T) & = \Omega([\rho_0,s_0],\mu,T) \\
   & = E[\rho_0,s_0] - T\int d\rv s_0(\rv) + \int d\rv\rho_0(\rv)(v(\rv)-\mu),
\end{align}
and that the internal energy in equilibrium is
\begin{align}
  E_0(N_0,S_0) = E[\rho_0,s_0],
\end{align}
where $S_0 = \int d\rv s_0(\rv)$.

Using the definition (\ref{EQomegaDefinition}) the Euler-Lagrange
equations (\ref{EQomegaDerivativesVanish}) can be cast in the form
\begin{align}
  \left.\frac{\delta E[\rho,s]}{\delta \rho(\rv)}\right|_{\rho_0,s_0} &= \mu - v(\rv),
  \label{EQELrho}\\
  \left.\frac{\delta E[\rho,s]}{\delta s(\rv)}\right|_{\rho_0,s_0} &= T.
  \label{EQELs}
\end{align}
For completeness, the Helmholtz free energy functional, on which DFT
is conventionally built, is obtained as
\begin{align}
  F([\rho],T) &= \min_s \left(E[\rho,s] - T\int d\rv s(\rv) \right)
  \label{EQhelmholtzAsMinimization}\\
   &= E[\rho,s_\rho] - T \int d\rv s_\rho(\rv),
\end{align}
where $s_\rho(\rv)$ denotes the entropy distribution at the minimum in
(\ref{EQhelmholtzAsMinimization}), which hence satisfies 
\begin{align}
  \left.\frac{\delta E[\rho,s]}{\delta s(\rv)}\right|_{\rho,s_\rho}=T,
\end{align}
where $\rho(\rv)$ is the (trial) density distribution on the left hand
side of (\ref{EQhelmholtzAsMinimization}).

Eqs.~(\ref{EQELrho}) and (\ref{EQELs}) constitute a closed system of
equations for the determination of $\rho_0(\rv)$ and $s_0(\rv)$ for
given thermodynamic statepoint $\mu,T$ and given external potential
$v(\rv)$. In practical applications one is required to use an
approximation for $E[\rho,s]$. Hence it is interesting to formulate
common free energy DFT approximations in the internal energy picture,
as we do in the next section.

\section{Examples for Internal Energy Functionals}
\label{SECapproximations}
We start with the ideal gas, where $U(\rv_1,\ldots,\rv_N)=0$. The
Helmholtz free energy functional can be derived from the fact that the
absence of interactions decouples all volume elements of the system
\cite{evans79,hansen06}. In each volume element the (bulk) ideal gas
properties holds. Hence the free energy functional is an integral over
a local free energy density,
\begin{align}
  F_{\rm id}([\rho],T) = k_BT \int d\rv \rho(\rv) 
  \left[\ln(\rho(\rv)\Lambda^3)-1\right],
  \label{EQFid}
\end{align}
where the thermal de Broglie wavelength depends on $T$ and is given by
\begin{align}
  \Lambda = \sqrt{\frac{2\pi\hbar^2}{mk_BT}},
\end{align}
and $\hbar=h/(2\pi)$. Corresponding reasoning leads to the internal
energy functional either by starting directly from the expression for
the bulk internal energy of the ideal gas, or by Legendre transforming
the integrand in (\ref{EQFid}). One arrives at the result
\begin{align}
  E_{\rm id}[\rho,s] = \frac{3\pi\hbar^{2}}{{\rm e}^{5/3}m}
  \int d\rv \rho(\rv)^{5/3} 
  \exp\left(\frac{2s(\rv)}{3k_B\rho(\rv)}\right),
  \label{EQEid}
\end{align}
where ${\rm e}$ is the exponential constant. This result is the same
as that obtained from Legendre transforming each volume element. The
functional (\ref{EQEid}) is local and non-linear. Note that
$\hbar^2/m$ carries units of ${\rm energy}\times {\rm length}^2$, as
is consistent with the integrand that has units of ${\rm
  length}^{-5}$. Clearly, the comparison of (\ref{EQEid}) to
(\ref{EQFid}) points to the quite striking density power of $5/3$ in
(\ref{EQEid}), and the fact that the entropy density per unit volume,
$s(\rv)$, appears in effect as an entropy density per particle,
$s(\rv)/\rho(\rv)$.

Evaluating the derivatives in the Euler-Lagrange equations
(\ref{EQELrho}) and (\ref{EQELs}) and rearranging yields
\begin{align}
  \rho_0(\rv) &= \Lambda^{-3}\exp\left(-\frac{\mu-v(\rv)}{k_BT}\right),\\
  s_0(\rv) &= -k_B \rho_0(\rv)\left[ \ln(\rho_0(\rv)\Lambda^3) - 5/2 \right],
\end{align}
and insertion into (\ref{EQEid}) gives the internal energy of the
ideal gas, solely due to kinetic contributions, $E_{\rm
  id}[\rho_0,s_0]= 3k_BT\int d\rv \rho(\rv)/2$, a result which is
certainly as expected.

For interacting systems the total Helmholtz free energy is usually
split into an ideal and an excess (over ideal) contribution as
\begin{align}
  F([\rho],T)=F_{\rm id}([\rho],T)+F_{\rm exc}([\rho],T),
\end{align}
where $F_{\rm id}([\rho],T)$ is given by (\ref{EQFid}) and $F_{\rm
  exc}([\rho],T)$ describes the effects of interparticle
interactions. For hard spheres, most approximate functionals can be
written in the form
\begin{align}
  F_{\rm exc}([\rho],T) &= k_BT \int d\rv \Phi(\rv),
  \label{EQFhardSpheres}
\end{align}
where $\Phi(\rv)$ is a scaled excess free energy density per unit
volume, which is independent of $T$. Temperature enters only via the
global scaling factor $k_BT$. For non-local functionals $\Phi(\rv)$ is
a functional of $\rho(\rv)$, typically via convolution. When such
additional convolution integrals are present in the functional the
choice which integral features as the ``outer'' integral in
(\ref{EQFhardSpheres}) is not necessarily unique; see
appendix~\ref{SECappendixFMT} for a discussion of a suitable choice in
fundamental-measure theory ~\cite{rosenfeld89,kierlik90}. Consider the
following form of the internal energy functional
\begin{align}
  E_{\rm HS}[\rho,s] = \frac{3\pi\hbar^2}{{\rm e}^{5/3}m}
  \int d\rv \rho(\rv)^{5/3} \exp\left(
  \frac{s(\rv)-s_{\rm HS}([\rho],\rv)}{3k_B\rho(\rv)/2}
  \right),
  \label{EQEhardSpheres}
\end{align}
where $s_{\rm HS}([\rho],\rv)=-k_B\Phi(\rv)$ is the hard sphere
contribution to the entropy. Eq.~(\ref{EQEhardSpheres}) is equivalent
to (\ref{EQFhardSpheres}) as can be seen from evaluating the
Euler-Lagrange equations (\ref{EQELrho}) and (\ref{EQELs}), which
yield
\begin{align}
  s_0(\rv) &= -k_B \rho_0(\rv)\left[\ln(\rho_0(\rv)\Lambda^3)-5/2\right] 
  +s_{\rm HS}([\rho_0],\rv)],\label{EQeulerLagrangeHSentropy}\\
  \rho_0(\rv) &= \Lambda^{-3} \exp\left(
  \frac{\mu-v(\rv)}{k_BT}+
  c_{\rm HS}^{(1)}([\rho_0],\rv)
  \right), \label{EQeulerLagrangeHSdensity}
\end{align}
where $c_{\rm HS}^{(1)}$ is equivalent to the one-body direct
correlation function for hard spheres and obtained here as
\begin{align}
  c_{\rm HS}^{(1)}([\rho],\rv) =
  k_B^{-1}\frac{\delta}{\delta \rho(\rv)}\int d\rv' s_{\rm HS}([\rho],\rv').
\end{align}

The common random phase approximation (RPA) \cite{hansen06} consists
of splitting a given interparticle pair interaction potential
$\phi(r)$, where~$r$ is the particle-particle distance and
$U(\rv_1,\ldots,\rv_N)=\sum_{i<j}\phi(|\rv_i-\rv_j|)$, into a
short-ranged repulsive, say hard sphere part $\phi_{\rm HS}(r)$ and a
long-ranged and slowly varying contribution $\phi_*(r)$, so that
$\phi(r)=\phi_{\rm HS}(r)+\phi_*(r)$. The corresponding internal
energy functional is
\begin{align}
  E_{\rm RPA}[\rho,s] &= E_{\rm HS}[\rho,s] + \frac{1}{2} \int d\rv \int d\rv'
  \rho(\rv)\rho(\rv') \phi_*(|\rv-\rv'|),
  \label{EQEmeanFieldAddition}
\end{align}
where the effects of $\phi_{\rm HS}(r)$ are described by the hard
sphere functional
(\ref{EQEhardSpheres}). Eq.~(\ref{EQEmeanFieldAddition}) leads to the
same ``entropic'' Euler-Lagrange equation
(\ref{EQeulerLagrangeHSentropy}) as for hard spheres, because $\delta
E_{\rm RPA}/\delta s(\rv) = \delta E_{\rm HS}/\delta s(\rv)$, and
generates an additional contribution $-\int
d\rv'\rho(\rv')\phi_*(|\rv-\rv'|)/(k_BT)$ inside of the exponential in
the ``density'' Euler-Lagrange equation
(\ref{EQeulerLagrangeHSdensity}).

For any system where the bulk internal energy $E_0(N_0,S,V)$ is known,
division by volume yields an internal energy density
$\epsilon_0(\rho,s)=E_0(N_0/V,S/V)/V$, from which in a local
density approximation (LDA) the functional
\begin{align}
  E_{\rm LDA}[\rho,s] = \int d\rv \epsilon_0(\rho(\rv),s(\rv))
  \label{EQElda}
\end{align}
follows. This is expected to be a good approximations when the
smallest length scale over which $\rho(\rv)$ and $s(\rv)$ vary is much
larger than all correlation lengths in the system.

A further ``generic'' approximation, analogous to the
Ramakrishnan-Youssouf (RY) \cite{ramakrishnan79} quadratic
approximation, is to truncate the functional Taylor expansion around a
homogeneous state with $\rho(\rv)=\rho_b=\rm const$ and
$s(\rv)=s_b=\rm const$ at second order in density,
\begin{widetext}
\begin{align}
  E_{\rm RY}[\rho,s] & = E_b(\rho_b,s_b)+\frac{1}{2}
  \int d\rv \int d\rv'
  \left[ \Delta \rho(\rv) \Delta \rho(\rv') \dE_{\rho\rho}(|\rv-\rv'|)
    \right.\nonumber\\ & \quad\quad
    \left.+ 2\Delta \rho(\rv) \Delta s(\rv') \dE_{\rho s}(|\rv-\rv'|)
    + \Delta s(\rv) \Delta s(\rv') \dE_{ss}(|\rv-\rv'|)
    \right],
  \label{EQEfunctionalExpansion}
\end{align}
\end{widetext}
where $\Delta\rho(\rv)=\rho(\rv)-\rho_b$ and $\Delta
s(\rv)=s(\rv)-s_b$ are the deviations from the respective bulk values,
the subscript $b$ indicates bulk quantities and the
$\dE_{\rho\rho}(r), \dE_{\rho s}(r)$ and $\dE_{ss}(r)$ are the second
functional derivatives of $E[\rho,s]$ evaluated in the homogeneous
bulk,
\begin{align}
  \dE_{ab}(|\rv-\rv'|) = \left.\frac{\delta^2 E[\rho,s]}{\delta a(\rv) \delta b(\rv')}
  \right|_{\rho_b,s_b},
  \quad a,b=\rho,s.
\end{align}
In general,
\begin{align}
  \dE_{\rho\rho}(\rv,\rv') &= 
  \left.\frac{\delta^2 E[\rho,s]}{\delta \rho(\rv) \delta \rho(\rv')}
  \right|_{\rho_0,s_0}, \label{EQcErhorho}\\
  \dE_{\rho s}(\rv,\rv') &= 
  \left.\frac{\delta^2 E[\rho,s]}{\delta \rho(\rv) \delta s(\rv')}
  \right|_{\rho_0,s_0},\\
  \dE_{ss}(\rv,\rv') &= 
  \left.\frac{\delta^2 E[\rho,s]}{\delta s(\rv) \delta s(\rv')}
  \right|_{\rho_0,s_0},
  \label{EQcEss}
\end{align}
Note that the first order terms in (\ref{EQEfunctionalExpansion})
vanish, as one expands around equilibrium and hence the Euler-Lagrange
equations (\ref{EQELrho}) and (\ref{EQELs}) hold (in the case
$v(\rv)=0$). 

Note that (\ref{EQcErhorho})-(\ref{EQcEss}) are analogous to the usual
two-body direct correlation function obtained from the excess
Helmholtz free energy functional as
\begin{align}
  c_2(\rv,\rv') = -(k_BT)^{-1}
  \left.
  \frac{\delta^2 F_{\rm exc}([\rho],T)}{\delta\rho(\rv)\delta\rho(\rv')}
  \right|_{\rho_0}
\end{align}

\section{Diffusive Dynamics}
\label{SECdynamics}
Using the equilibrium framework developed in
Sec.\ \ref{SECvariationalPrinciple}, we find it interesting to use it
in a dynamical context, similar in spirit to dynamical density
functional theory (DDFT) which rests on the equilibrium Helmholtz free
energy functional. Much current research activity is aimed at applying
and developing DDFT, which provides a dynamical equation for the time
evolution of the density profile $\rho(\rv,t)$, where $t$ is time. In
order to derive such an equation the continuity equation for the
density profile, which is exact, is supplemented by approximations for
the ``thermodynamic driving force'' that acts on the density. As
compared to a diffusion equation gradients in chemical potential are
replaced by gradients in the functional (density) derivative of the
Helmholtz free energy density.  Starting from a more microscopic point
of view, DDFT can also be derived from the Smoluchowski equation
\cite{archer04ddft}.

Here, we spell out a similar framework for the joint time evolution of
$\rho(\rv,t)$ and the time and position-dependent entropy distribution
$s(\rv,t)$. We keep the discussion at a phenomenological level and
make no attempts at a derivation from first principles, albeit paying
attention that fundamental symmetry relations, i.e., the Onsager
reciprocal relations, are satisfied. Hence the strategy consists of
taking the appropriate dynamic equations from linear irreversible
thermodynamics \cite{onsager31one,onsager31two} and replacing the
temperature and density fields in the continuum description by the
microscopic (functional) derivatives of the internal energy
functional.

We impose two continuity equations, one for the density $\rho(\rv,t)$
and one for the internal energy density $\epsilon(\rv,t)$,
\begin{align}
  \dot \rho(\rv,t) &= -\nabla\cdot\J_\rho(\rv,t),
  \label{EQdotDensity}\\
  \dot \epsilon(\rv,t) &= -\nabla\cdot\J_\epsilon(\rv,t),
  \label{EQdotEnergy}
\end{align}
where the dot denotes a partial time derivative,
i.e.\ $\dot\rho=\partial \rho/\partial t$ and $\dot\epsilon=\partial
\epsilon/\partial t$. Solving the Gibbs-Duhem relation $d\epsilon =
Tds + \mu d\rho$ for the differential entropy per unit volume, $ds$,
gives
\begin{align}
  ds &= \frac{1}{T}d\epsilon - \frac{\mu}{T}d\rho,
  \label{EQgibbsDuhem}
\end{align}
from which the prefactors of the differentials on the right hand side
are identified as the ``driving forces'' for the internal energy
current $\J_\epsilon$ and for the particle density current
$\J_\rho$. Hence
\begin{align}
  \J_\rho &= D \rho \nabla \frac{-\mu}{k_BT}
  +D_T \epsilon \nabla \frac{1}{k_BT}
  \label{EQcurrentRho}
  ,\\
  \J_\epsilon &= D_T \epsilon \nabla \frac{-\mu}{k_BT}
  +D_{\rm th} \frac{\epsilon^2}{\rho} \nabla \frac{1}{k_BT},
  \label{EQcurrentEpsilon}
\end{align}
where we have omitted the arguments $\rv, t$, and have introduced the
particle diffusion coefficient $D$, the thermal diffusion coefficient
$D_T$, and the thermal conductivity $D_{\rm th}$, all of which possess
dimensions of ${\rm length}^2/{\rm time}$.  The powers of $\rho$ and
$\epsilon$ in the prefactors of the gradients in (\ref{EQcurrentRho})
and (\ref{EQcurrentEpsilon}) can be determined from dimensional
analysis, by observing that the left hand side of (\ref{EQcurrentRho})
possesses units of $1/({\rm time}\times\rm{length}^2)$ and that of
(\ref{EQcurrentEpsilon}) has units of ${\rm energy}/({\rm
  time}\times{\rm length}^2)$.  Note that the ``cross terms'',
i.e.\ the prefactor of the second gradient in (\ref{EQcurrentRho}) and
of the first gradient in (\ref{EQcurrentEpsilon}) are identical as
requested by the Onsager reciprocal relations. See appendix
\ref{SECdissipationFunction} for a derivation of (\ref{EQcurrentRho})
and (\ref{EQcurrentEpsilon}) starting from a dissipation function.
The change in entropy is obtained via the Gibbs-Duhem relation
(\ref{EQgibbsDuhem}) as
\begin{align}
  \dot s & = \frac{1}{T} \dot \epsilon - \frac{\mu}{T} \dot \rho\\
  &= -\frac{1}{T} \nabla\cdot\J_\epsilon + \frac{\mu}{T} \nabla\cdot\J_\rho,
  \label{EQdotEntropy}
\end{align}
where (\ref{EQdotEntropy}) follows from the continuity equations
(\ref{EQdotDensity}) and (\ref{EQdotEnergy}).

Bearing in mind the structure of the Euler-Lagrange equations
(\ref{EQELrho}) and (\ref{EQELs}), we replace $\mu$ by $\delta
E[\rho,s]/\delta \rho(\rv,t) + v(\rv,t)$ and $T$ by $\delta
E[\rho,s]/\delta s(\rv,t)$. Here we have allowed the external
potential to be time-dependent, in order to model a corresponding
external influence on the system. Hence we rewrite
(\ref{EQdotDensity}) and (\ref{EQdotEnergy}) as
\begin{align}
  \J_\rho
  &= -\frac{D}{k_B}\rho\nabla \frac{\dE_\rho + v}{\dE_s}
    +\frac{D_T}{k_B} \epsilon \nabla \frac{1}{\dE_s},
  \label{EQDynamicalEDFrho}\\
  \J_\epsilon
 &= -\frac{D_T}{k_B}\epsilon\nabla \frac{\dE_\rho + v}{\dE_s}
    +\frac{D_{\rm th}}{k_B} \frac{\epsilon^2}{\rho} \nabla \frac{1}{\dE_s},
  \label{EQDynamicalEDFs}
\end{align}
where we
have used the short-hand notation for the first functional derivatives
of the internal energy functional,
\begin{align}
  \dE_\rho &= \left.\frac{\delta E[\rho,s]}{\delta \rho(\rv)}
  \right|_{\rho(\rv,t),s(\rv,t)},\\
  \dE_s &= \left.\frac{\delta E[\rho,s]}{\delta s(\rv)}
  \right|_{\rho(\rv,t),s(\rv,t)}.
\end{align}
Performing the replacement of the local temperature and the local
chemical potential by the corresponding functional derivatives in
(\ref{EQdotEntropy}) yields
\begin{align}
  \dot s = -\frac{1}{\dE_s}\nabla\cdot\J_\epsilon
  +\frac{\dE_\rho +v}{\dE_s}\nabla\cdot\J_\rho.
  \label{EQdotsWithFunctionalDerivatives}
\end{align}
The equations for the currents (\ref{EQDynamicalEDFrho}) and
(\ref{EQDynamicalEDFs}) together with the continuity equation for the
particle density (\ref{EQdotDensity}) and for the energy density
(\ref{EQdotEnergy}), along with
(\ref{EQdotsWithFunctionalDerivatives}), form a closed set of
equations for the time evolution of $\rho(\rv,t)$ and $s(\rv,t)$, for
given $v(\rv,t)$ and initial conditions $\rho(\rv,0)$ and $s(\rv,0)$
at time $t=0$. In general the diffusion coefficients $D, D_T$ and
$D_{\rm th}$ will depend on $\rho(\rv,t)$ and $s(\rv,t)$; assuming
them to be constant would be the simplest approximation.

\section{Conclusions}
\label{SECconclusions}

In summary, we have developed a variational formulation of classical
statistical mechanics, which is centered around the internal energy as
a functional of the one-body density distribution $\rho(\rv)$ and the
position-dependent entropy distribution $s(\rv)$. Although the
definition of $s(\rv)$ is not unique [cf.\ Eq.~(\ref{EQs0}) for the
  equilibrium value $s_0(\rv)$], the current choice possesses two
important properties that make it a suitable variable in the
variational framework: i) the space integral of $s(\rv)$ is the
macroscopic entropy, and ii) the definition is local in the sense that
it probes the entropy under the condition that a particle resides at
the space point $\rv$ considered. One of the Euler-Lagrange equations
for the minimization of the grand potential is very similar to that of
DFT based on the Helmholtz theory, i.e., the functional derivative
with respect to the density field is related to a local chemical
potential, cf.~(\ref{EQELrho}). Physically, such a situation can be
realized by an external potential acting on the system.  The internal
energy functional $E[\rho,s]$ depends on the local density $\rho(\rv)$
and on the entropy distribution $s(\rv)$. The functional derivative
with respect to $s(\rv)$ gives the (constant) temperature in
equilibrium, cf.\ (\ref{EQELs}). Having this further Euler-Lagrange
equation is to be considered a strength of the theory, when it comes
to applications using an approximate functional. Rather than having to
implement the physics of $T=\rm const$ on the level of the
approximation for the free energy functional, the internal energy
functional offers an additional mechanism to relax to equilibrium via
an inhomogeneous entropy distribution.

In Levy's constrained search method, which we used for formulating the
variational framework, there is no need for introducing a field that
is conjugate to the local entropy distribution. Hence the situation is
different from the local chemical potential that is conjugate to the
one-body density. The relationship between these thermodynamic fields
plays a crucial role in the standard Mermin-Evans formulation of
DFT. However, in equilibrium there is at least no simple conjugate to
the entropy distribution. Such a role would be played by an
position-dependent temperature, which we deliberately avoided in the
derivation presented in Sec.~\ref{SECvariationalPrinciple}.

Obtaining dynamical equations for the density and entropy
distributions is straightforward when using linear irreversible
thermodynamics in a continuum description as a starting point and
replacing the fields for temperature and chemical potential by the
appropriate functional derivatives of the internal energy functional,
cf.\ (\ref{EQcurrentRho}) and (\ref{EQcurrentEpsilon}). This approach
is phenomenological and we have made no attempts at deriving the
dynamics from first principles under controlled approximations for the
microscopic dynamics. While the structure of the dynamic equations is
a straightforward extension of dynamical DFT, there is also an
important distinction: When using the Helmholtz free energy
functional, in principle any (non-pathological) density field is a
physically realizable one via choice of an appropriate external
potential. The situation is different when considering the internal
energy functional and prescribing both the density field and the
entropy field. In general, no corresponding equilibrium situation will
exist, i.e., one cannot choose an external potential and a
temperature, cf.\ (\ref{EQELrho}) and (\ref{EQELs}), so that the given
trial fields $\rho(\rv)$ and $s(\rv)$ become equilibrium
quantities. This effect is far less subtle than that of
representability of trial density fields, cf.\ the discussion in
\cite{dwandaru11levy}. 

Clearly, true hydrodynamic effects, that originate from local momentum
and angular momentum conservation, are neglected in the treatment of
Sec.~\ref{SECdynamics}. However, there remains a wide range of
interesting physics associated entirely with diffusive dynamics in
(complex) liquids \cite{kohler02,wiegand04}, see e.g.\ Dhont's
treatment of thermodiffusion \cite{dhont04one,dhont04two}.

We have formulated a variety of standard DFT approximations in
internal energy language. The mathematical structure of some of these
functionals appear unfamiliar in a variational context, cf.\ the form
of the ideal gas internal energy functional and the way in which the
ideal gas and interaction contributions are coupled in the case of
hard spheres, cf.~(\ref{EQEhardSpheres}). Other approximations are
consistent with expectation, i.e., the addition of a mean-field energy
contribution (\ref{EQEmeanFieldAddition}), the local density
approximation (\ref{EQElda}) and the Taylor expansion up to second
order around a homogeneous (fluid) state
(\ref{EQEfunctionalExpansion}).

The potential importance of the current work lies i) in the additional
insights that can be gained from studying the entropy distribution in
applications within existing approximations such as these described in
Sec.\ \ref{SECapproximations}, and ii) in the possibility to construct
internal energy functionals that couple the density and entropy
contributions in novel ways. Investigating the implications for the
dynamical test particle limit \cite{archer07dtpl,hopkins10dtpl} is an
interesting topic for future work, as is considering quenched-annealed
mixtures \cite{schmidt02pordf,schmidt02aom,lafuente06qadft} and the
dynamics of atomic liquids \cite{archer06atomic}. Finally note that
changing the thermodynamical potential as we have done here is very
different from changing to a different ensemble, see
e.g.\ \cite{White00prl} for DFT in the canonical ensemble.

{\em Note added in proof.--} The current theory possess similarities,
but also significant differences to the approach by Phil Attard
\cite{attard}. His theory is, broadly speaking, based on the entropy
functional with the internal energy being a variable.

\hspace{3mm} \acknowledgments I thank R. Evans for a critical reading
of the manuscript, and H.\ R.\ Brand, W.\ K\"ohler, and
Th.\ M.\ Fischer for useful discussions. This work was supported by
the SFB840/A3 of the DFG.

\appendix
\section{Fundamental-measure theory}
\label{SECappendixFMT}

Although Rosenfeld's functional [for a hard sphere mixture with
  one-body density profile $\rho_i(\rv)$ of species $i$] possesses the
structure of (\ref{EQFhardSpheres}),
\begin{align}
  F_{\rm exc}[\{\rho_i\}] = \int d\xv \Phi(\xv),
\end{align}
the position coordinate $\xv$ is very different from the argument
$\rv$ of the entropy distribution (\ref{EQfTOs}). Rather than
corresponding to a particle position, $\xv$ is a mere convolution
integral that couples the FMT weight functions in order to represent
the hard sphere Mayer bond, and for third and higher orders in
density, constitutes the center of star diagrams that are formed by
weight function bonds \cite{leithall11hys}.

Both the definition of the entropy field (\ref{EQfTOs}), and in the
ratio of entropy and density distribution in the exponential of the
hard sphere internal energy functional (\ref{EQEhardSpheres}), one
expects a particle to be located at the position considered. In FMT
there is no shortage of position integrals over the density, hence the
problem is to single out one of them in a non-biased, ``symmetric''
way.

In order to achieve this we start from the power series of FMT
\cite{leithall11hys}, which reads
\begin{align}
  \F_{\rm exc}[\{\rho_i\}] = 
    k_BT \int d\xv \sum_{m=2}^\infty\frac{1}{m(m-1)}[\N(\xv)]^m
  \label{EQfmtAsPowerSeries}
\end{align}
where $[\N(\xv)]^m$ is the $m$-th matrix power of
\begin{align}
  \N(\xv) &=  \left(\begin{matrix}
    n_3(\xv) & n_2(\xv) & n_1(\xv) & n_0(\xv)\\
    0 & n_3(\xv) & \frac{n_2(\xv)}{4\pi} & n_1(\xv)\\
    0 & 0 & n_3(\xv) & n_2(\xv)\\
    0 & 0 & 0 & n_3(\xv)
  \end{matrix}\right).
\end{align}
Here the weighted densities $n_\alpha(\xv)$ are obtained by
convolution,
\begin{align}
  n_\alpha(\xv) = \sum_i \int d\rv \rho_i(\rv) w_\alpha^{(i)}(\rv-\xv),
\end{align}
where the $w_\alpha(\cdot)$ are the Kierlik-Rosinberg FMT weight
functions \cite{kierlik90}, and the sum is over all hard sphere
species~$i$. Note that $\F_{\rm exc}[\{\rho_i\}]$ in
(\ref{EQfmtAsPowerSeries}) is a $4\times 4$-matrix and that the
physically relevant entry $F_{\rm exc}[\{\rho_i\}]$ is that in the
first row and last column \cite{leithall11hys}.

We rewrite the integrand in (\ref{EQfmtAsPowerSeries}) as
\begin{align}
 &  \N(\xv) \sum_{m=2}^\infty \frac{1}{m(m-1)}[\N(\xv)]^{m-1}\\
 & = \sum_i \int d\rv {\sf W}^{(i)}(\xv-\rv) \rho_i(\rv)
  \sum_{m=2}^\infty \frac{1}{m(m-1)}[\N(\xv)]^{m-1},
  \label{EQfmtIntegrand}
\end{align}
where the matrix of weight functions is defined as
\begin{align}
  {\sf W}^{(i)}(\xv) &=  \left(\begin{matrix}
    w_3^{(i)}(\xv) & w_2^{(i)}(\xv) & w_1^{(i)}(\xv) & w_0^{(i)}(\xv)\\
    0 & w_3^{(i)}(\xv) & \frac{w_2^{(i)}(\xv)}{4\pi} & w_1^{(i)}(\xv)\\
    0 & 0 & w_3^{(i)}(\xv) & w_2^{(i)}(\xv)\\
    0 & 0 & 0 & w_3^{(i)}(\xv)
  \end{matrix}\right).
\end{align}
Reintroducing the $\xv$-integral and re-arranging in
(\ref{EQfmtIntegrand}) gives
\begin{align}
  \F_{\rm exc}[\{\rho_i\}] &= k_BT \int d\rv \sum_i \rho_i(\rv) \nonumber\\
  & \qquad\times \int d\xv {\sf W}_i(\xv-\rv)
  \sum_{m=2}^\infty \frac{1}{m(m-1)}[\N(\xv)]^{m-1}, \\
  & \equiv
  k_BT \int d\rv \sum_i \rho_i(\rv) \Psi_i(\rv),
  \label{EQfmtIntegrandTwo}
\end{align}
where we have defined the free energy density (per particle) for
species $i$ as
\begin{align}
  \Psi_i(\rv) = \int d\xv {\sf W}_i(\xv-\rv)
  \sum_{m=2}^\infty \frac{1}{m(m-1)}[\N(\xv)]^{m-1}.
\end{align}
Hence we single out the entry in the top right corner of the matrix in
(\ref{EQfmtIntegrandTwo}), and rewrite it as
\begin{align}
  F_{\rm exc}[\{\rho_i\}] = \int d\rv \sum_i \rho_i(\rv) 
    \sum_{\alpha=0}^3 \left(w_\alpha^{(i)} \ast \phi_\alpha\right)(\rv),
    \label{EQfexcFinal}
\end{align}
where the asterisk denotes the convolution, and explicit
expressions for the $\phi_\alpha(\xv)$ are as follows:
\begin{align}
  \phi_0 &= 1+\left(\frac{1}{n_3}-1\right)\ln(1-n_3),\\
  \phi_1 &= - \frac{n_2}{n_3} - \frac{n_2}{n_3^2}\ln(1-n_3),\\
  \phi_2 &=  \left(\frac{n_2^2}{4\pi n_3^3}-\frac{n_1}{n_3^2}\right)\ln(1-n_3)
              -\frac{n_1}{n_3} + 
              \frac{n_2^2(2-n_3)}{8\pi n_3^2(1-n_3)},\\
  \phi_3 &= -\left (\frac{n_0}{n_3^2}-\frac{2n_1n_2}{n_3^3}+\frac{n_2^3}{4\pi n_3^4}
           \right)\ln(1-n_3) \nonumber\\\
         &\qquad - \frac{n_0}{n_3}
           +\frac{n_1 n_2 (2-n_3)}{n_3^2(1-n_3)}
           -\frac{n_2^3(2n_3^2-9n_3+6)}{24\pi n_3^3(1-n_3)^2}.
\end{align}
In summary, the integrand in (\ref{EQfexcFinal}) forms a
suitable choice for the desired quantity, i.e.,
\begin{align}
  s_{\rm HS}([\{\rho_i\}],\rv) & =-k_B\Phi(\rv) \\
  & = -k_B \sum_i\rho_i(\rv) \sum_{\alpha=0}^3
  \left( w_\alpha^{(i)} \ast \phi_\alpha  \right)(\rv).
\end{align}

\section{Dissipation function}
\label{SECdissipationFunction}
As a consistency check on (\ref{EQcurrentRho}) and
(\ref{EQcurrentEpsilon}) [and hence (\ref{EQDynamicalEDFrho}) and
  (\ref{EQDynamicalEDFs})], we derive the currents from a (scaled)
dissipation function $R$ \cite{onsager31one,onsager31two}, which we
assume to be given by
\begin{align}
  R &= \frac{D\rho}{2}\left(\nabla\frac{-\mu}{k_BT}\right)^2
     +D_T \epsilon\left(\nabla\frac{-\mu}{k_BT}\right)\cdot
           \left(\nabla\frac{1}{k_BT}\right)
  \nonumber\\ & \qquad\qquad
     + \frac{D_{\rm th}\epsilon^2}{2\rho}
           \left(\nabla\frac{1}{k_BT}\right)^2.
\label{EQdissipationFunction}
\end{align}
Here $R$ is a scaled object with dimensions of $({\rm
  length}\times{\rm time})^{-1}$.  One can verify explicitly that
(\ref{EQdissipationFunction}) generates the expressions
(\ref{EQcurrentRho}) and (\ref{EQcurrentEpsilon}) via
\begin{align}
 \J_\rho = \frac{\partial R}{\partial \left(\nabla\frac{-\mu}{k_BT}\right)},\\
 \J_\epsilon = \frac{\partial R}{\partial\left(\nabla\frac{1}{k_BT}\right)}.
\end{align}
Furthermore, one can show explicitly that for the entropy production
\begin{align}
  \dot s + \nabla\cdot\J_s = 2k_BR,
\end{align}
holds, where the entropy current is
\begin{align}
  \J_s = \frac{1}{T} \J_\epsilon -\frac{\mu}{T} \J_\rho,
\end{align}
consistent with the Gibbs-Duhem relation (\ref{EQgibbsDuhem}).

Finally note that (\ref{EQcurrentRho}) and (\ref{EQcurrentEpsilon})
can be written as a matrix product
\begin{align}
  \left(\begin{matrix}
    \J_\rho\\\\
    \J_\epsilon
  \end{matrix}\right) =
  \left(\begin{matrix}
    D \rho\quad & D_T \epsilon\\\\
    D_T \epsilon\quad & D_{\rm th} \epsilon^2/\rho
   \end{matrix} \right)\cdot
  \left(\begin{matrix}
    \nabla\frac{-\mu}{k_BT}\\\\
    \nabla\frac{1}{k_BT}
   \end{matrix} \right),
  \label{EQcurrentsAsMatrix}
\end{align}
where the matrix of kinetic coefficients [on the right hand side of
  (\ref{EQcurrentsAsMatrix})] is symmetric, as requested by the
Onsager reciprocal relations \cite{onsager31one,onsager31two}.

\end{document}